# Self-focused pulse propagation is mediated by spatiotemporal optical vortices

M. S. Le[1,2], G. A. Hine[3], A. Goffin[1,4], J.P. Palastro[5], and H. M. Milchberg[1,2,4]

[1]*Institute for Research in Electronics and Applied Physics, University of Maryland, College Park, Maryland 20742, USA*
[2]*Dept. of Physics, University of Maryland, College Park, Maryland 20742, USA*
[3]*Oak Ridge National Laboratory, Oak Ridge, Tennessee 37831, USA*
[4]*Dept. of Electrical and Computer Engineering, University of Maryland, College Park, Maryland 20742, USA*
[5]*University of Rochester, Laboratory for Laser Energetics, Rochester, New York, 14623 USA*
[4]*milch@umd.edu*

We show that the dynamics of high-intensity laser pulses undergoing self-focused propagation in a nonlinear medium can be understood in terms of the topological constraints imposed by the formation and evolution of spatiotemporal optical vortices (STOVs). STOVs are born from point phase defects on the sides of the pulse nucleated by spatiotemporal phase shear. These defects grow into closed loops of spatiotemporal vorticity that initially exclude the pulse propagation axis, but then reconnect to form a pair of toroidal vortex rings that wrap around it. STOVs constrain the intrapulse flow of electromagnetic energy, controlling the focusing-defocusing cycles and pulse splitting inherent to nonlinear pulse propagation. We illustrate this in two widely studied but very different regimes,  relativistic self-focusing in plasma and non-relativistic self-focusing in gas, demonstrating that STOVs mediate nonlinear propagation irrespective of the detailed physics.

Vortices are flow structures describing the circulation of a physical quantity about a local axis. They are ubiquitous in physics, appearing in sound waves and common fluids [1], superfluid helium [2,3], Bose-Einstein condensates [4], and aerodynamics [5]. They are intimately connected to orbital angular momentum (OAM) -carrying beams of photons or massive particles [6–9]. All of these vortices can be mathematically described using fixed spatial axes [6-11]. By contrast, the circulation associated with electromagnetic spatiotemporal optical vortices (STOVs) occurs in spacetime, where the local vortex axis is embedded in a propagating pulse and can be oriented orthogonal to propagation. STOVs were first measured as toroidal electromagnetic structures that naturally emerge from nonlinear self-focusing collapse and self-guiding of ultrashort laser pulses in air [12]. STOV pulses linearly propagating in free space were later generated and measured in [13-15], with a theory of transverse OAM presented in [16] and experimentally verified in [17].

The emergence of STOVs from self-focusing occurs in tandem with a collapse arrest mechanism [12]: a response of the medium that prevents unlimited growth of laser intensity and leads to self-guiding. As unlimited self-focusing will always be halted by *some* medium response, it is reasonable to expect that STOVs are a universal feature of any transient self-focusing process.

In this paper we show that, once formed, STOVs are topologically constrained to mediate intrapulse energy flows that determine global pulse evolution during self-guiding such as focusing and defocusing cycles and pulse splitting. STOVs nucleate from point phase defects or dislocations created by the extreme phase shear produced during self-focusing arrest. The dislocations grow as twisted spacetime vortex loops that exclude the propagation axis until they reconnect to form stable vortex rings that wrap around the propagation axis. STOV dynamics provide a unifying picture for nonlinear self-focusing processes independent of the detailed physics. This is illustrated by two cases with distinctly different physics, where the peak laser intensities are ~6 orders of magnitude apart: (1) filamentation of relativistic intensity pulses ($a_0 > 1$) in plasmas [18-21] and (2) filamentation of non-relativistic pulses ($a_0 \ll 1$) in gases [22,23]. Here $a_0 = eA_0/mc^2$ is the normalized vector potential, with $e$ is the electron charge, $A_0$ the peak laser vector potential, $m$ the electron mass, and $c$ the speed of light in vacuum.

During nonlinear propagation, STOV formation is initiated when the nonlinear phase shift at the center of a self-focusing beam strongly increases relative to phase at the beam periphery. This sets up a highly transient "boundary" in spacetime between high phase shift and low phase shift regions. Collapse arrest sufficiently stabilizes this boundary to promote nucleation of point phase defects, leading to the birth of STOVs. In the relativistic regime in plasma, the intensity-dependent refractive index of free electrons leads to beam collapse reaching intensities $>10^{18}$ W/cm$^2$, whereupon electrons near the beam axis are radially expelled by the laser ponderomotive force $\mathbf{F}_p = -mc^2\nabla(\gamma)$, where $\gamma = (1 + a_0^2/2)^{1/2}$ is the laser cycle-averaged Lorentz factor. This arrests collapse and sufficiently stabilizes the phase shift boundary that continued propagation accumulates $\pi$ phase shear, leading to nucleation of a point phase dislocation or defect (accompanied by a null in the field amplitude) followed by launching of STOVs. In non-relativistic filamentation in air, where beam collapse is caused by the nonlinear optical response of air molecules [24], collapse is arrested when intensities reach $> 10^{13}$ W/cm$^2$ and air is ionized [12]. Just as in the relativistic case, collapse arrest enables accumulation of $\pi$ phase shear and generation of STOVs.

We first discuss the role of STOVs in relativistically self-focused propagation in plasmas. Here, the laser fields are strongly nonperturbative and ponderomotive forces significantly distort the propagation medium. Under such conditions, perturbation-based envelope approaches such as the nonlinear Schrödinger equation [25] break down, and one uses particle-in-cell (PIC) codes which solve for the self-consistent laser oscillation-resolved electric and magnetic fields and the electron positions and momenta [26,27]. Ion motion is typically negligible during the ultrashort pulse interaction and is neglected. Details of the PIC simulations, plasma parameters, and variable definitions are presented in Appendix A.

To provide physical insight, we start with a cylindrically symmetric [26] simulation, where the initial phase defect and STOVs are constrained to emerge as rings around the propagation axis. Results are plotted in Fig. 1, where $\xi = v_g t - z$ is a local space coordinate in the moving computation window (the "beam frame") and $v_g = c(1 - \omega_p^2/\omega^2)^{1/2}$ is the laser group velocity in the plasma. In each frame, $\xi = 0$ references the initial pulse centroid moving at velocity $v_g$. Figures 1(a,a',a") show the pulse just before STOV generation, with plots of (a) isosurfaces $|E_x/E_0| = 0.1$ ($E_0$ is the peak electric field in each frame); (a') local electromagnetic phase circulation (colormap) and energy density flow (black arrows, see discussion below); and (a") laser phase fronts corresponding to (a'). As propagation proceeds, strong radial phase shear accumulates until it reaches $\pi$, nucleating a phase defect, shown as a black ring in Fig 1(b), a black dot at $\xi_{nuc} = 0.87\lambda_p$, $y_{nuc} = 0.59\lambda_p$ in the $\xi y$ plane of Fig. 1(b'), and as locally broken phase fronts in Fig. 1(b"). With further propagation, the radial phase jump exceeds $\pi$ and the wavefronts form a double forked structure (Fig. 1(c"), corresponding to the spawning of two vortex rings of opposite winding (or topological charge) wrapped around the pulse, with the blue ring leading the red ring (Fig 1(c)), also shown as blue and red dots in Fig. 1(c') centered on the vortex singularities. These toroidal vortex structures are STOVs. With continued pulse propagation, the STOVs separate: the leading

STOV advances while the trailing one falls behind, as seen in the progression from Fig. 1(c,c') to (d,d'). Figure 1(e), corresponding to Fig. 1(d, d'), plots the laser phase fronts and refractive index change $\Delta n = -\gamma^{-1} N_e/2N_{cr}$, from the ponderomotive force driving a plasma bubble of reduced electron density. Figure 1(d') is a blowup of the small white box in 1(e). Topological charge is conserved, so a STOV does not disappear except by collisional annihilation with another STOV of opposite winding.

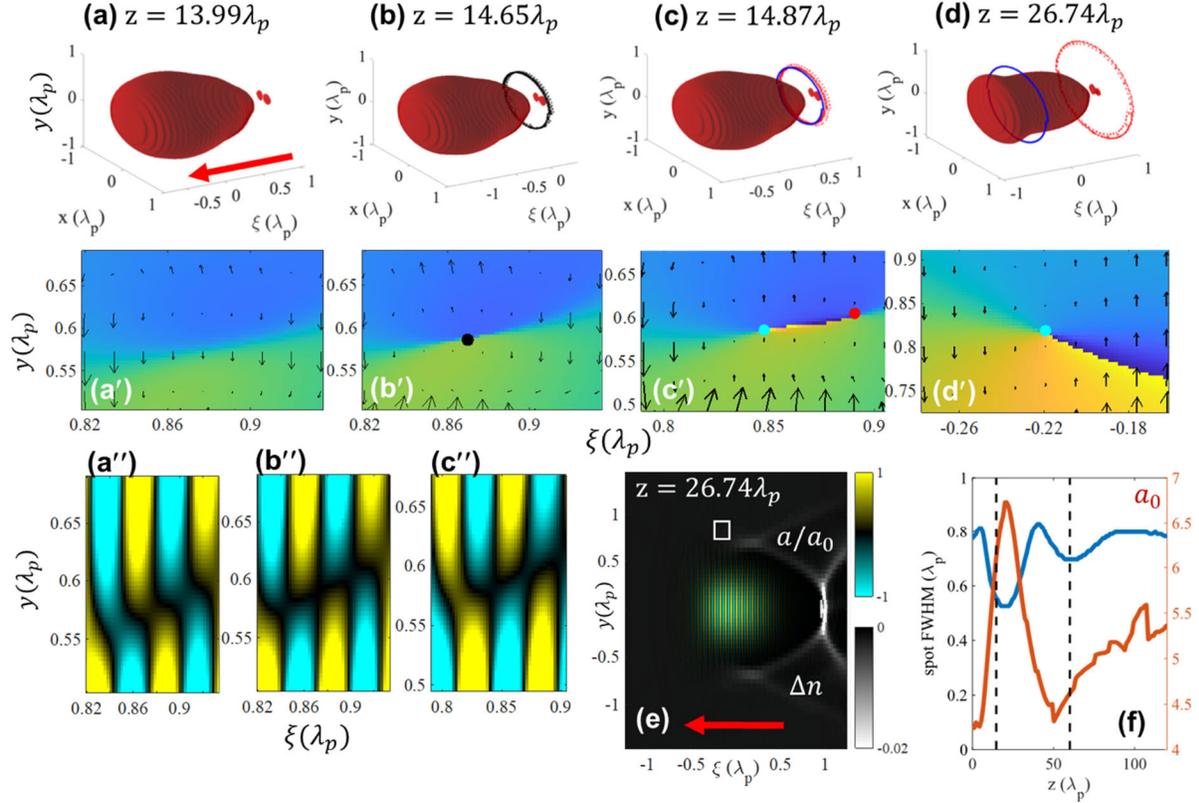

**Figure 1.** Particle in cell simulation (using FBPIC [26], described in Appendix A) of STOV formation by a $\tau = 30$ fs full-width-at-half-maximum laser pulse with $a_0 = 4.3$ and $1/e^2$ intensity spot size $w_0 = 9$ µm propagated into a semi-infinite fully ionized hydrogen plasma with $N_{e0} = 6 \times 10^{18}$ cm$^{-3}$ and $\lambda_p = 13.6$ µm. The beam waist is at the plasma entrance $z = 0$ and is matched for self-guiding (see text). Panels (a)-(d): $|E_x/E_0| = 0.1$ isosurface plots of laser pulse (red arrow is propagation direction) and loci of phase defect points (see Appendix C). Panels (a')-(d'): Electromagnetic phase circulation about the incipient phase defect ((a)) or spawning of STOVs in (b)-(d). Overlaid black arrows are proportional to the beam frame Poynting flux $\langle \mathbf{S}_{BF} \rangle$. Panels (a'')-(c''): pulse phase fronts corresponding to (a')-(c'), which show phase fronts almost breaking (a''), breaking and defect creation (b''), and reattachment to form a double forked pattern associated with $l = \pm 1$ vortices (c''). Panel (e): Plot of laser pulse phase fronts $a/a_0$ overlaid by refractive index shift of plasma $\Delta n = -\gamma^{-1} N_e/2N_{cr}$ (at $z = 26.74 \lambda_p$, corresponding to (d) and (d')). The small white box is frame (d'), centered on the singularity of the leading STOV (blue dot). Red arrow: pulse propagation direction. Panel (f): Pulse peak $a_0$ and spot FWHM vs. propagation distance. The first vertical dashed line marks $l = \pm 1$ STOV generation; the second line marks generation of a second STOV pair.

To plot the energy density flow in Fig. 1 (a'-d'), we calculated the optical-cycle-average beam frame Poynting vector $\langle \mathbf{S}_{BF} \rangle$, as described in Appendix B. The in-plane components of $\langle \mathbf{S}_{BF} \rangle$ are overlaid in Fig. 1(a')-(d') as scaled black arrows. The beam-frame energy flux is predominantly transverse with respect to the beam axis, consistent with the near light speed of the beam frame.

In Fig. 1(c')-(d'), energy density flux circulates around the STOV singularities (whose loci are found as described in Appendix C), with the leading STOV directing energy density radially inward (toward the propagation axis) in advance of the singularity and radially outward past the singularity. This vortex has winding number $l \equiv \Gamma/2\pi = 1$ for counterclockwise integration over a closed curve $C$ enclosing the vortex singularity in the upper $\xi y$ plane (see Appendices B and C). The trailing STOV, with $l = -1$, directs energy outward in advance of its singularity and inward past it. The entire region between the two STOVs is thus topologically constrained to convey pulse energy away from the propagation axis.

Figure 1(f) shows the large-scale effect of STOV evolution on pulse dynamics. First, as the pulse relativistically self-focuses, the peak field quickly increases from its initial value of $a_0 = 4.3$, while the pulse FWHM spot size drops from $\sim 0.8\lambda_p$. The black vertical dashed line at $z = 15\lambda_p$ indicates the onset of $l = \pm 1$ STOV generation from spatiotemporal phase shear. As the $l = 1$ and $l = -1$ STOVs move apart between Fig. 1(c) and 1(d), the increasing outward energy flow between the STOVs arrests self-focusing, and the pulse reaches its peak field and minimum spot size at $z \sim 20\lambda_p$. Propagating beyond this point, the reduction in local energy density by STOV-directed flow weakens self-focusing so that the spot size increases and $a_0$ drops. By then, the $l = 1$ STOV has moved to the front of the pulse and the $l = -1$ STOV remains at the back. In effect, global pulse dynamics, here shown as a cycle of focusing and defocusing, is determined by the intrapulse energy flow mediated by STOV evolution. With further propagation, the STOV singularities move radially away from the main pulse and their effect diminishes; continued phase shear causes wavefronts to curve inward again, the self-focusing process restarts past $z \sim 40\lambda_p$ and the spot size shrinks. In this second cycle, $a_0$ continues to increase from pulse compression in the plasma wake [18]. With sufficient spatiotemporal phase shear, a new pair of STOVs nucleates at $z \sim 60\lambda_p$ (at the second dashed line in Fig. 1(f)), whereupon the spot size begins to increase again. Note that pulse compression continues, accounting for the slow increase in $a_0$ past $z \sim 50\lambda_p$. By $z \sim 100\lambda_p$ the pulse becomes highly temporally modulated and breaks up from red shifting and dispersion.

In the axisymmetric simulations of Fig. 1, the initial phase defect and resultant STOVs were constrained to emerge as rings wrapped around the pulse. However, with the laser pulse linear polarization breaking this symmetry, the birth and initial evolution of STOVs is more complicated. This is shown in Fig. 2, which plots the evolution of STOVs determined from 3D PIC simulations [27] using the same laser and plasma parameters. First, point defects emerge on the top and bottom of the pulse (along $y$), perpendicular to the $x$ polarization axis (Fig. 2(a)). These point defects quickly evolve into expanding STOV loops on the top and bottom of the pulse (Fig. 2(b)). Because a loop can be spanned by an arbitrary surface that remains simply connected, it is topologically equivalent to a point, which is the limit of a shrinking loop [28]. As an aid to following the loop evolution in Fig. 2, loop segments that will eventually belong to the leading (trailing) STOV are blue (red). With continued propagation, the top and bottom loops expand and reconnect to each other on the far side of the plot to form a larger loop. Vortex reconnection is "catalyzed" by the appearance of smaller additional loops (Fig. 2(c)-(e), shown in black) which grow from point defects on the far side. Each expanding loop encloses outward energy flow, so adjacent segments of different loops have opposed phase windings. As they expand into each other (Fig. 2(e) and (f)), these segments annihilate, leading to formation of one large loop. A similar process is catalyzed on the near side (Fig. 2(f)-(h)), leaving two toroidal STOVs wrapped around the pulse (Fig. 2(h) and (i)). The evolution from point defect nucleation to oppositely charged STOVs takes $\sim 10$ wavelengths of propagation, or $\sim 30$ fs. The leading and trailing STOVs are similar to those in Fig.

1(c), where they are close together at the back of the pulse (see [29]). They then separate, driving the global pulse dynamics. An estimate for the rate of longitudinal vortex separation is $d\Delta\xi_{STOV}/dt \sim v_g\lambda/2\pi R$ [12,29], where $\Delta\xi_{STOV}$ is the separation between the leading and trailing toroidal STOVs and $R$ is the STOV ring radius. The $1/R$ dependence is consistent with vortex ring longitudinal motion in fluids [1]. This gives an approximate separation timescale of $\Delta\tau_{STOV}\sim(2\pi R/\lambda)\tau \sim 1.6$ ps, using an average $R\sim7$ μm from Fig. 1, consistent with the separation time of ~1.2 ps from the plot.

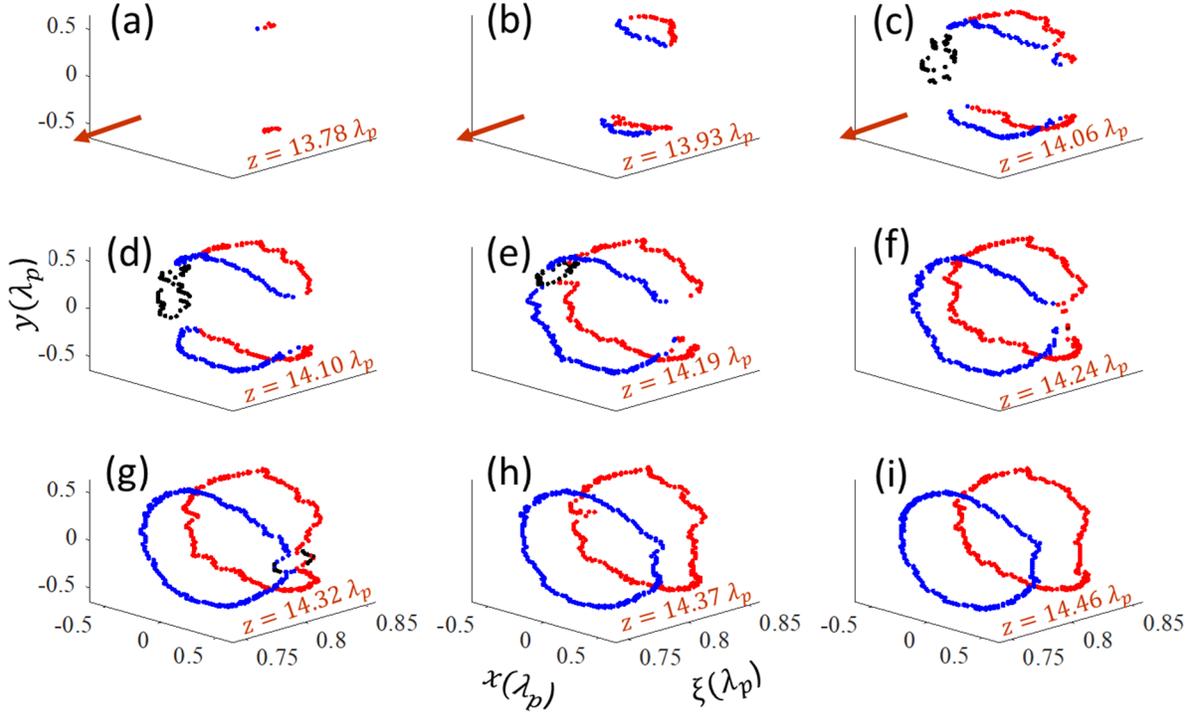

**Figure 2.** 3D PIC simulations (using EPOCH [27], described in Appendix A) of the birth and evolution of STOVs in relativistic self-guiding, for the laser and plasma conditions of Fig. 1, where $\lambda_p = 13.6$ μm is the plasma wavelength and the laser pulse is linearly polarized along $x$. The pulse propagation direction is indicated by the red arrows. STOVs grow from initial nucleation of point phase defects ((a)) to closed loops excluding the propagation axis ((b)-(e)), to a sequence of loop reconnection events ((d)-(h)) that lead to a pair of ring STOVs of opposite winding wrapped around the propagation axis ((h) and (i)). The loci of phase defects are determined from the PIC code output as described in Appendix C. As an aid to following the loop evolution, loop segments that will eventually belong to the leading (trailing) STOV are blue (red), and the catalysis loop(s) are black.

An interesting effect is the location dependence of the point defect nucleation on the initial field amplitude $a_0$. At $a_0 = 4.3$, the $x$-polarized field drives electrons in the beam periphery to higher average momentum in the $x\xi$ plane than in the $y\xi$ plane, relatively increasing the local $\gamma$ in the $x\xi$ plane and reducing the local plasma refractive index contribution $-\gamma^{-1}N_e/2N_{cr}$. This decreases spatiotemporal phase shear along $x$ in favor of along $y$. As $a_0$ decreases or increases from this range, variation of $N_e$ in the polarization direction plays a greater relative role in phase shear, with defects first appearing along $x$. Our simulations show that any source of azimuthal asymmetry, such as polarization and beam shape, can lead to localized generation of point defects and ultimately, via vortex reconnection, to $l = \pm 1$ toroidal STOVs wrapping around the pulse. The evolution timescale varies with the strength of the asymmetry, an example of which is shown in [29] for a pulse with an elliptical spot injected into the plasma.

We now examine the role of STOVs in non-relativistic filamentation in gases. We use our 3D unidirectional pulse propagation code YAPPE [30,31] to simulate self-focusing and filamentation in atmospheric pressure argon of 25 mJ, 500 fs FWHM pulses with $P/P_{cr} = 5$, where $P_{cr}$ is the critical power for self-focusing. Argon is chosen for its near-instantaneous electronic nonlinearity, which avoids STOV generation by delayed transient effects such as molecular rotation in air [12]. The linearly polarized Gaussian beam is launched at its waist ($w_0 = 1$ mm) into a semi-infinite gas slab. We choose a longer pulse than typically used in filamentation experiments to extend the length of high intensity filamentary propagation, as shown in [31,32], here to ~0.5 m. Unlike in the relativistic case of Figs. 1 and 2, there is no light-pressure-induced distortion of the propagation medium.

Figure 3 shows the pulse evolution, where $z$ is distance from the front of the slab and the pulse propagates right to left in each panel. Figure 3(a) ($z = 70$ cm) shows the pulse in the process of collapsing, where the inward pointing white arrows are proportional to the beam-frame Poynting vector $\langle \mathbf{S}_{BF} \rangle$, calculated (see Appendix B) using $\widetilde{\mathbf{D}}(\mathbf{r}_\perp, \omega; z) = \varepsilon(\mathbf{r}_\perp, \omega; z) \, \widetilde{\mathbf{E}}(\mathbf{r}_\perp, \omega; z)$, where the dielectric function $\varepsilon$ includes the neutral argon and plasma response [30,31], which is first calculated in the time domain and then Fourier transformed. Collapse continues through $z = 98$ cm until arrested when the pulse intensity exceeds the ionization threshold for argon. Coincident with collapse arrest, spatiotemporal phase shear produces a defect which generates $l = \pm 1$ toroidal STOVs illustrated as blue ($l = +1$) and red ($l = -1$) dots in the $x\xi$ plane projections in Fig. 3(c)-(h). In these simulations, which assume a uniform medium and no field-driven asymmetry, the STOVs emerge as rings wrapped around the pulse. Simulating self-focusing of the same pulse in turbulent argon gas [29] shows STOVs originating from point defects, as seen in Fig. 3(i)-(l), and similar to Fig. 2. The evolution from point defects to ring STOVs takes ~2.7 ps of propagation.

As the pulse propagates beyond $z{\sim}105$ cm, the STOVs move apart and the entire region between them is topologically constrained to direct energy radially outward, similar to the relativistic case of Fig. 1. However, defocusing and refocusing manifests differently than in the relativistic case, where laser energy is radially confined in its self-generated plasma density depression. Figure 3(c)-(h) shows that as propagation proceeds, outward energy flow on the STOVs' near sides, combined with inward energy flow on their far sides, leads to spatiotemporal focusing of energy on the far side of each STOV, forming two peaks. This is pulse-splitting, a well-known feature of non-relativistic self-focusing [33]. In prior work, pulse-splitting had been qualitatively explained as the nonlinear refocusing of the low intensity, off-axis pulse energy 'reservoir' surrounding the on-axis intensity peak, or 'core' [34]. Filamentation and pulse splitting have also been interpreted in terms of "X-wave" propagation [35-37]. The X-shaped intensity structures shown in Fig. 3(c)-(h) are very similar to those previously discussed [35-37]. Here, we see that the presence of STOV cores and their accompanying field nulls is responsible for the X-shapes.

For most of this evolution, the leading STOV advances and slightly lags the leading portion of the pulse, directing energy toward the axis, while the trailing STOV lags the back of the pulse, directing that part of the energy outward and beginning another pulse defocusing and focusing cycle, manifested here as another cycle of pulse splitting. The transition from single pulse to double pulse in Fig. 3 takes ~0.9 ns of propagation, consistent with the estimate $\Delta\tau_{STOV}{\sim}(2\pi R/\lambda)\tau \sim 0.8$ ns for $R{\sim}200$ μm, the mean ring radius of the $l = \pm 1$ STOVs.

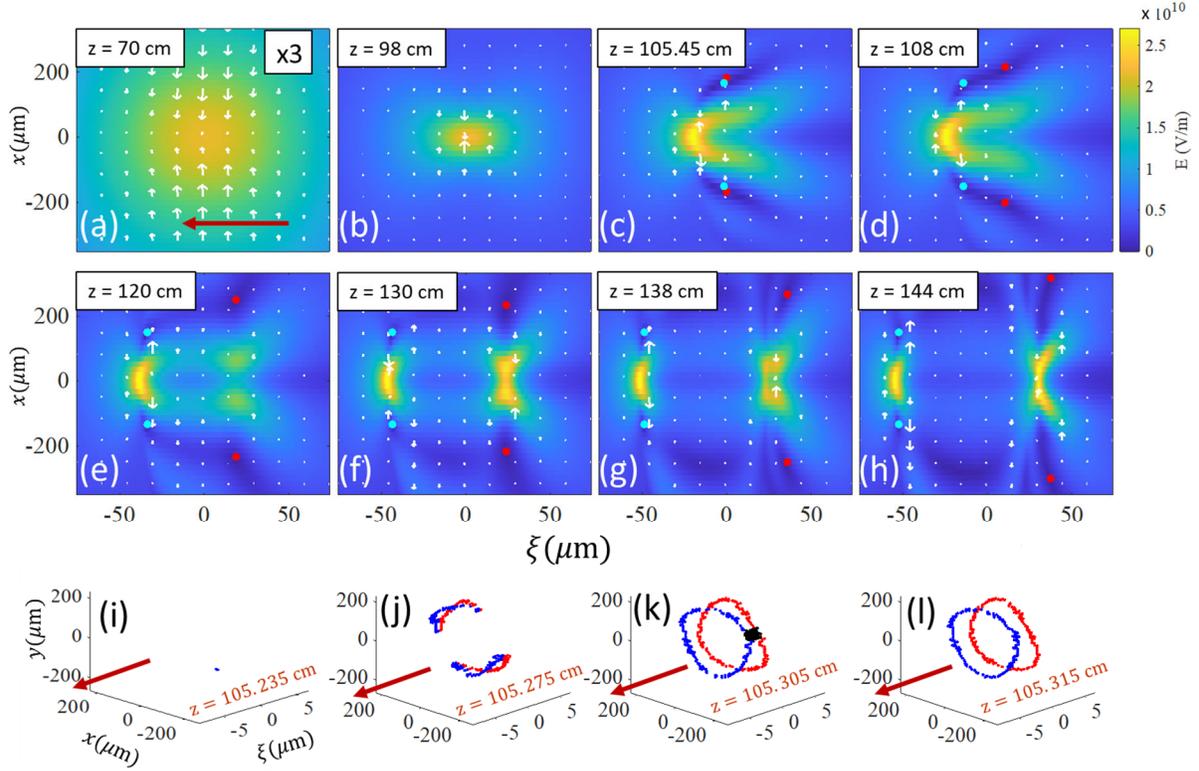

**Figure 3.** Simulation (YAPPE [31,32]) of a nonrelativistic Gaussian laser pulse (25 mJ, 500 fs FWHM, $\lambda = 0.8$ μm, $w_0 = 1$ mm, $P/P_{cr} = 5$ ) is launched into a semi-infinite argon gas of density $2.5 \times 10^{19}$ cm$^{-3}$. The red arrows indicate the propagation direction. The beam waist is located at the gas entrance at $z = 0$. The laser envelope, the transverse beam frame Poynting vector $\langle \mathbf{S}_{BF} \rangle$ (proportional to white arrows), and STOVs (leading/lagging as blue/red dots) are plotted. Prior to STOV formation ((a) (intensity × 3) and (b)), energy flow is primarily inward, indicating self-focusing. Just before panel (c), a defect has spawned $l = \pm 1$ STOVs. As the two STOVs move apart ((c) though (h)), energy is topologically constrained to flow outward from the propagation axis everywhere between them, but inward on their far sides, leading to pulse splitting. (i)-(l): Simulation in turbulent argon [29] showing point defect nucleation and STOV generation. Underline{Simulation parameters}: Window: 0.5 cm ($x$) × 0.5 cm ($y$) × 0.06 cm ($\xi$); grid dimensions: $\Delta x = \Delta y = 16.67$ μm, $\Delta \xi = 0.75$ μm.

In summary, global pulse dynamics in intense self-focused propagation can be understood as a consequence of topological constraints imposed by STOVs. STOVs nucleate from point defects induced by pulse asymmetry-favoured spatiotemporal phase shear on the sides of the pulse, expand into loops excluding the propagation axis, and undergo reconnection to form oppositely charged vortex rings wrapping around the propagation axis. These ring STOVs separate with propagation, topologically mandating outward flow of electromagnetic energy between the STOVs (on their near sides) and inward flow on their far sides. In relativistic filamentation in plasma, this process drives the transition from self-focusing to defocusing and back. For non-relativistic filamentation in gas, STOV pair evolution drives pulse splitting. STOV dynamics provide a unifying picture that explains global pulse evolution for widely different laser intensities and propagation medium responses.

Finally, one might ask whether STOVs are causal agents or mere spectators. The answer is that STOVs are self-consistent structures that both result from and drive the energy flow. Consider the analogy of vortices in fluids, which arise from shear induced in the fluid velocity field. Once formed, fluid vortices cannot in any sense be considered non-perturbative spectators to the global flow process—they have clearly become determinative. In this respect, flows of electromagnetic energy in a light pulse are no less fluid-like.

The authors thank Scott Hancock, Lucas Railing, Sina Zahedpour, and Nihal Jhajj for technical discussions. This work is supported by the US Dept. of Energy (DESC0024406), National Science Foundation (PHY2010511), and the Air Force Office of Scientific Research (FA9550-21-1-0405).

*Appendix A: Particle-in-cell (PIC) simulations*– Relativistic simulations of this paper were performed using the cylindrical PIC code FBPIC [26] and a fully 3D+1 PIC code EPOCH [27] for solving the Maxwell-Lorentz system of equations. A $\tau = 30$ fs FWHM $x$-linearly polarized laser pulse propagating along $z$ was focused to a $1/e^2$ intensity beam waist $w_0 = 9$ μm (vacuum Rayleigh range $z_0 = 31$ μm) and $a_0 = 4.3$ into a semi-infinite fully ionized hydrogen plasma with electron density $N_{e0} = 6 \times 10^{18}$ cm$^{-3}$, with the beam waist located at the vacuum-plasma interface. Here $N_{e0}/N_{cr} = 0.0035$, where $N_{cr} = m\omega^2/4\pi e^2 = 1.7 \times 10^{21}$ cm$^{-3}$ is the critical density at the laser central wavelength $\lambda = 2\pi c/\omega = 810$ nm, and $m$ and $e$ the electron rest mass and charge. These conditions satisfy the matched spot size condition for self-guiding, $k_p w_0 \cong 2\sqrt{a_0}$ [21], where $k_p = \omega_p/c$ is the plasma wavenumber ($\lambda_p = 2\pi/k_p = 13.6$ μm is the plasma wavelength) and $\omega_p = (4\pi e^2 N_{e0}/m_e)^{1/2}$ is the plasma frequency.

*Appendix B: Beam frame Poynting flux*– The beam frame Poynting flux is $\langle \mathbf{S}_{\mathrm{BF}} \rangle = (\langle S_x \rangle, \langle S_y \rangle, v_g \langle u \rangle - \langle S_z \rangle)$, where $\langle \mathbf{S} \rangle = (c/8\pi) \boldsymbol{\mathcal{E}} \times \boldsymbol{\mathcal{H}}^*$ and $\langle u \rangle = (16\pi)^{-1}(\boldsymbol{\mathcal{E}} \cdot \boldsymbol{\mathcal{D}}^* + \boldsymbol{\mathcal{B}} \cdot \boldsymbol{\mathcal{H}}^*)$ are the cycle averaged Poynting vector and electromagnetic energy density. The script quantities $\boldsymbol{\mathcal{E}}$, $\boldsymbol{\mathcal{D}}$, $\boldsymbol{\mathcal{B}}$, $\boldsymbol{\mathcal{H}}$ are the complex spatiotemporal field envelopes, where in the plasma we take $\boldsymbol{\mathcal{H}} = \boldsymbol{\mathcal{B}}$ and $\boldsymbol{\mathcal{D}} = \boldsymbol{\mathcal{E}} + i(4\pi/\omega) \boldsymbol{j}$, where $\boldsymbol{j}$ is the complex spatiotemporal current density envelope. The script quantities are calculated from the real valued PIC code-generated fields, say $\mathbf{E}(\mathbf{r}_\perp, \xi; z)$, as $\boldsymbol{\mathcal{E}}(\mathbf{r}_\perp, \xi; z) = (2\pi)^{-1} \int_0^\infty d\omega\, \widetilde{\mathbf{E}}(\mathbf{r}_\perp, \omega; z) e^{-i\omega\tau}$ (eliminating negative $\omega$), where $\widetilde{\mathbf{E}}(\mathbf{r}_\perp, \omega; z) = \int_{-\infty}^\infty d\tau\, \mathbf{E}(\mathbf{r}_\perp, \xi; z) e^{i\omega\tau}$ is the time Fourier transform of $\mathbf{E}$, $\mathbf{r}_\perp = (x, y)$, $\tau = \xi/v_g$, and propagation distance $z$ plays the role of a running parameter.

*Appendix C: Phase singularity detection*– The loci of phase singularities (the STOV cores) in this paper were determined by numerically evaluating the spatiotemporal phase circulation $\Gamma = \oint_C \nabla\Phi \cdot d\boldsymbol{l}$ over three contours $C$ in three intersecting planes ($xy$, $y\xi$, and $\xi x$) around each grid point in the simulation, where $\Phi(\mathbf{r}_\perp, \xi; z) = \arg(\boldsymbol{\mathcal{E}})$ and $\nabla = (\partial/\partial x, \partial/\partial y, \partial/\partial\xi)$. If any of the three integrals were within 2% of $2\pi$, the point within the contour was taken to be a phase defect and added to a locus of similar points. With respect to each of the 3 planes, each point is at the common vertex of four rectangular grid cells, and $C$ follows the perimeter of the group of four cells. We note that $C$, here set by the grid, can in principle be arbitrarily small and is not restricted in size by the laser wavelength [38].

# Supplementary Material: Self-focused pulse propagation is mediated by spatiotemporal optical vortices

M. S. Le[1,2], G. A. Hine[1,2,3], A. Goffin[1,4], J.P. Palastro[5], and H. M. Milchberg[1,2,4,*]

[1]Institute for Research in Electronics and Applied Physics, University of Maryland, College Park, Maryland 20742, USA

[2]Dept. of Physics, University of Maryland, College Park, Maryland 20742, USA

[3]Oak Ridge National Laboratory, Oak Ridge, Tennessee 37831, USA

[4]Dept. of Electrical and Computer Engineering, University of Maryland, College Park, Maryland 20742, USA

[5]University of Rochester, Laboratory for Laser Energetics, Rochester, New York, 14623 USA

[*]milch@umd.edu

## 1. Additional simulations

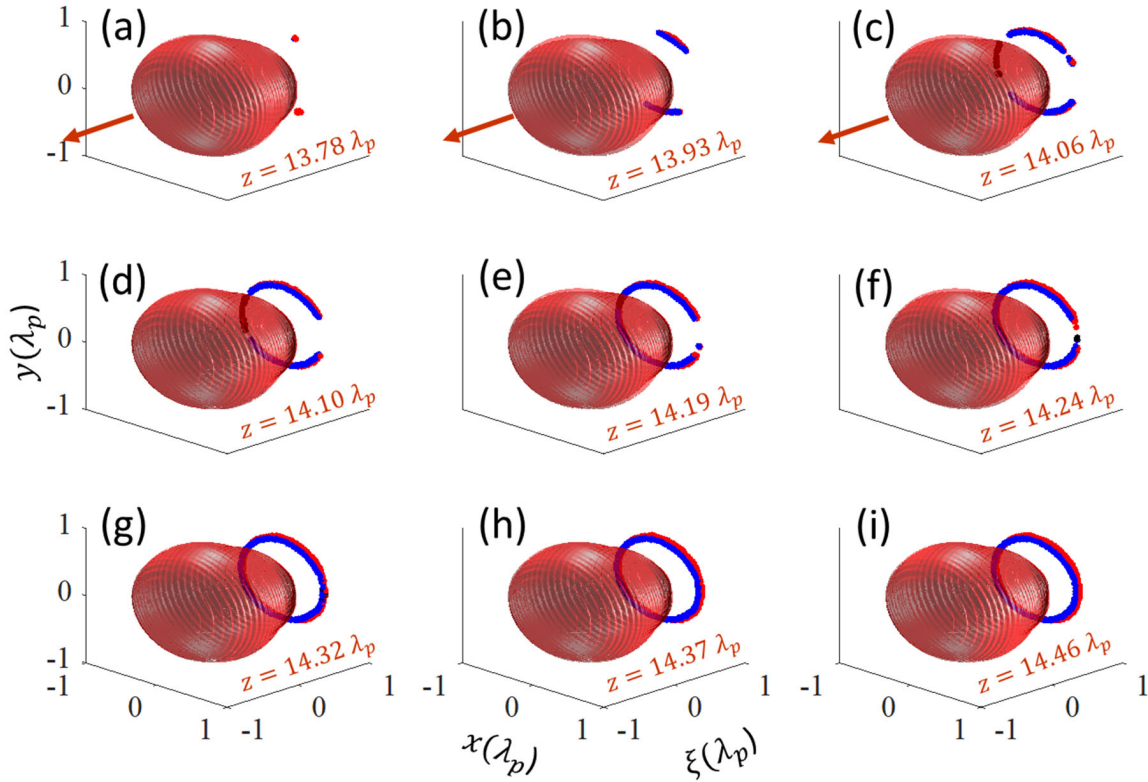

**Figure S1.** 3D+1 PIC simulation (using EPOCH [1]) corresponding to Fig. 2 of main paper, where the range of $\xi$ is expanded to show the whole laser pulse and the relative location of initial point defect generation (a), vortex loop growth and reconnection ((b)-(g)), and launching of toroidal STOVs that wrap around the pulse ((h),(i))). Here is plotted the $|E_x/E_0| = 0.1$ isosurfaces of the laser pulse (red arrow is propagation direction) and the loci of phase defect points in blue (red) that will be part of the leading (trailing) STOV, and catalytic loops in black. Here $\lambda_p = 13.6 \ \mu m$.

Figure S1 serves as a companion to Fig. 2 of the main paper, putting the rapid formation of STOVs into the context of a wider view of the laser pulse. The evolution from point defects to vortex loops excluding the propagation axis, followed by reconnection to form two STOVs of opposite winding wrapping around the pulse axis, occurs over ~30 fs of propagation. This entire process takes place in a longitudinally narrow window to the rear of the pulse. After formation, the ring STOVs look irregular on the magnified longitudinal scale of Fig. 2, but they appear far more symmetric at-scale in Fig. S1.

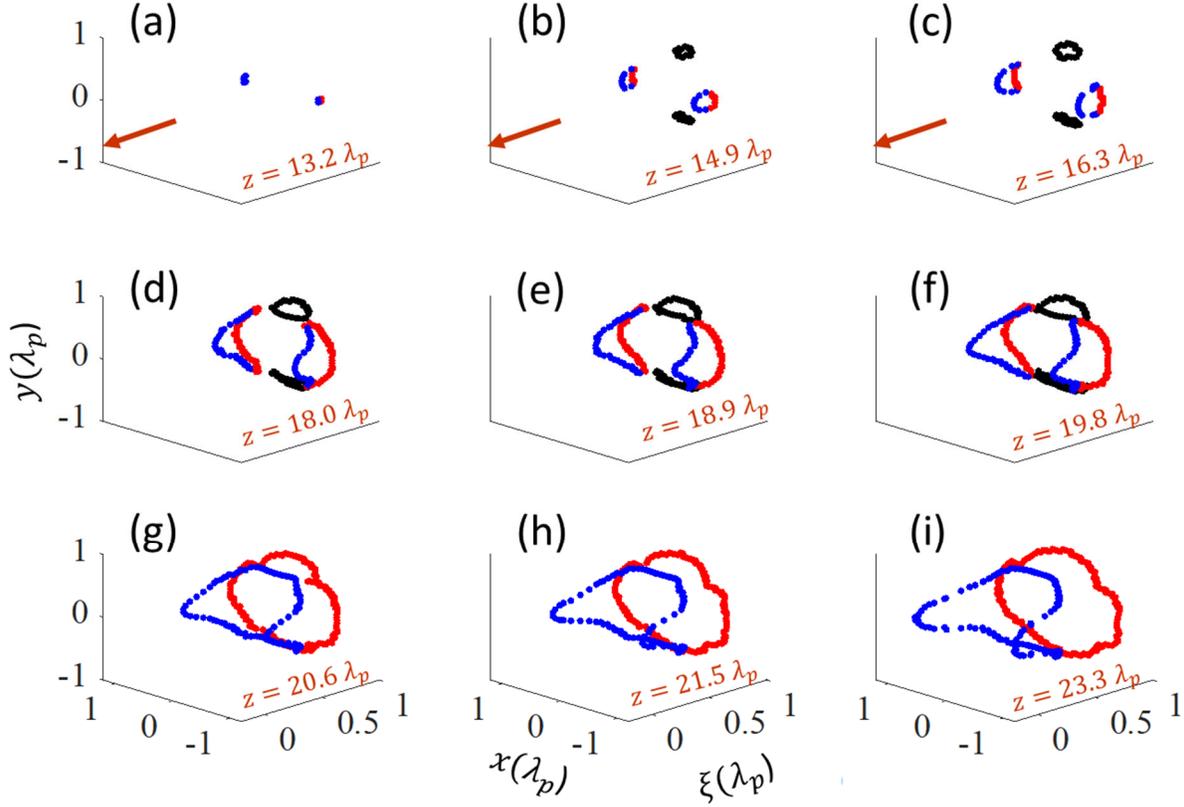

**Figure S2.** EPOCH 3D+1 simulation of an elliptical beam waist with $w_{0y}/w_{0x} = 2$, where $k_p\sqrt{w_{0y}w_{0x}} \cong 2\sqrt{a_0}$ for the same plasma conditions of Figs. 1−3, where $a_0 = 4.3$. The red arrows in the top row indicate the pulse propagation direction. As an aid to following the loop evolution, loop segments that will eventually belong to the leading (trailing) STOV are blue (red), and the catalysis loop(s) are black. The evolution from initial point defect (in (a)) to two ring STOVs wrapped around the propagation axis (panel (g)) takes ∼350 fs of propagation time, over $10\times$ longer than the symmetric case of Fig. 2. Also, note that the extent of the $\xi$ scale is comparable to those in Fig. 1 and Fig. S1, so that the leading (blue) STOV loop is axially stretched over a significant time interval of the pulse.

For spatially asymmetric pulse shapes, vortex evolution can occur over a longer time and result in distorted STOV shapes. Figure S2 shows a PIC simulation [1] of an $x$-linearly polarized pulse with elliptical beam waist $w_{0y}/w_{0x} = 2$, where we set $k_p\sqrt{w_{0y}w_{0x}} \cong 2\sqrt{a_0}$, keeping the same initial $a_0 = 4.3$ used in the simulations of Figs. 1−3. In (a), point defects appear at the sides of the pulse along $x$. Soon after ((b)-(f)), four STOV loops form (2 major loops along $x$ and 2 smaller "catalytic" loops along $y$, all excluding the propagation axis). These loops all enclose outward flowing energy density so that adjacent segments of nearby loops have opposite phase windings and annihilate as they expand into each other. After ∼350 fs of pulse propagation (> $10\times$ that in Fig. 2), the reconnection process is over by panel (g), where two STOVs wrap around the propagation axis. The leading STOV, whose $x$-sides are stretched forward at this stage of propagation, reflects the defocusing in $x$ and self-focusing in $y$ (greater outward energy flux through $y\xi$ planes than through $x\xi$ planes). The extent of the $\xi$ scale is comparable to those in Fig. 1 and Fig. S1, so that the sides of the leading (blue) STOV loop are axially stretched over much larger portion of the pulse compared to Fig. 2 (as seen in Fig. S3).

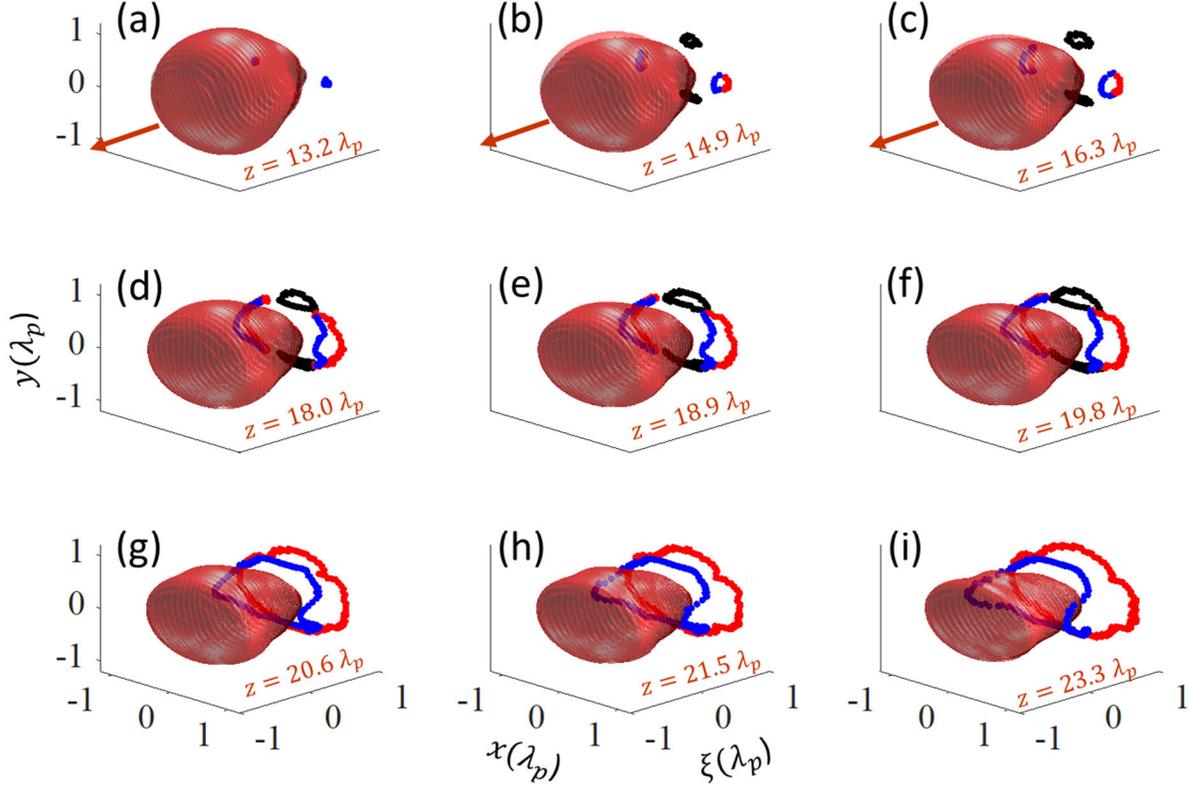

**Figure S3**. Companion figure to Fig. S2 for elliptical pulse simulation, where the range of $\xi$ is expanded to show the relative location, with respect to the laser pulse, of initial point defect generation (a), vortex loop growth and reconnection ((b)-(f)), and launching of toroidal STOVs that wrap around the pulse ((g)-(i)). The $|E_x/E_0| = 0.1$ isosurfaces of the laser pulse (red arrow is propagation direction) are plotted and the loci of phase defect points is shown with vortex segments in blue (red) that will be part of the leading (trailing) STOV, and catalytic loops in black. Here $\lambda_p = 13.6$ μm. Here, STOV evolution and reconnection takes place over a time scale $\sim 10 \times$ that in the symmetric case of Fig. 1.

The initial formation of point defects along $x$ is favored by the sharper intensity gradients in that direction from the smaller $w_{0x}$. Our simulations show that sufficient initial spatial asymmetry in the pulse can dominate the effects of polarization in seeding the initial point defects leading to STOV generation.

Figure S4 illustrates the asymmetric nucleation of STOVs during non-relativistic filamentation in argon at atmospheric density using 3D +1 YAPPE simulations [2,3]. Asymmetry was introduced into the simulations by adding turbulence via a modified von Karman turbulence model applied with phase screens [5], using an inner scale (smallest scale of refractive index variation) of 1 mm, an outer scale (largest scale) of 1 m, and $C_n^2 = 6.4 \times 10^{-14}$ m$^{-2/3}$. The first point defect emerges in panel (a) on the side of the pulse, followed by one on the other side. By panel (b), two loops have grown from the point defects and expand until reconnection takes place over panels (d) and (e), with similar catalysis loops (black points) as in Fig. 2 promoting the reconnection. By (e), there is a single large folded loop that wraps around the pulse. By (f), with the aid of a catalytic loop, the folded loop has connected with itself, producing the two familiar ring STOVs that move apart. The evolution from point defects to ring STOVs takes ~2.7 ps of propagation.

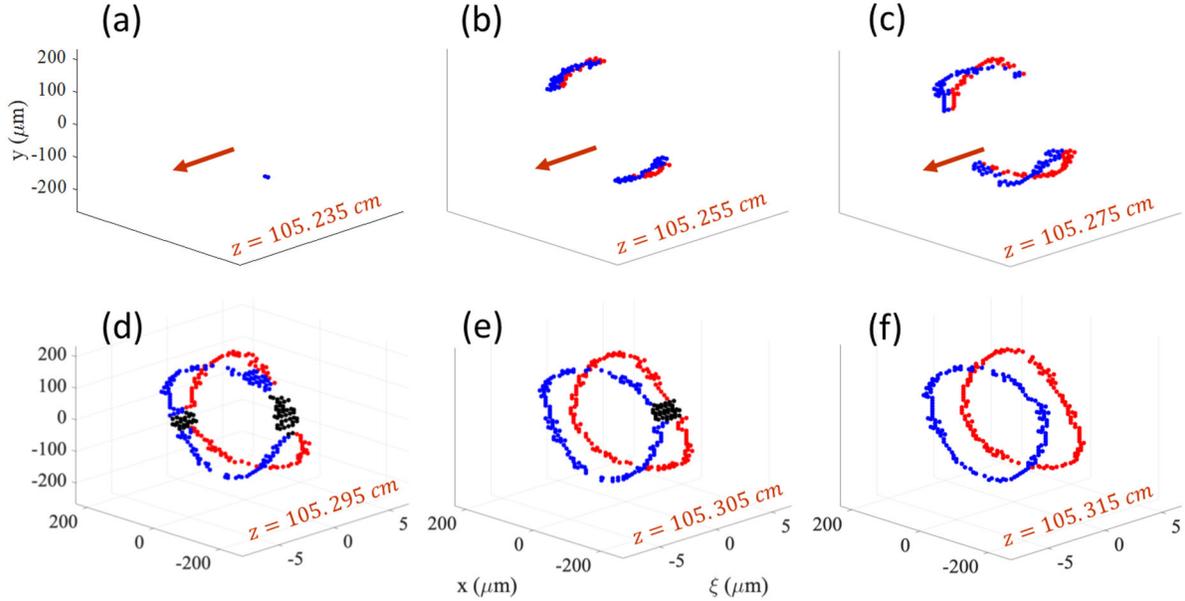

**Figure S4.** Simulation (YAPPE [2,3]) of a nonrelativistic Gaussian laser pulse (25 mJ, 500 fs FWHM, $\lambda = 0.8$ μm, $w_0 = 1$ mm, $P/P_{cr} = 5$ ) launched into a semi-infinite turbulent argon gas of average density $2.5 \times 10^{19}$ cm$^{-3}$. The beam waist $z = 0$ is at the gas entrance. Turbulence induces vortex loop growth from point defects, with eventual reconnection to form $l = \pm 1$ STOVs. The simulation used a modified von Karman turbulence model applied with phase screens [5], using an inner scale (smallest scale of refractive index variation) of 1 mm, an outer scale (largest scale) of 1 m, and $C_n^2 = 6.4 \times 10^{-14}$ m$^{-2/3}$.

## 2. Separation of $l = \pm 1$ STOVs

As calculated in ref. [4], the motion of the vortex core of a toroidal $l = \pm 1$ STOV is described by

$$k_0 \frac{\partial \mathbf{r}_{\text{STOV}}}{\partial z} = \pm \frac{1}{\rho} \left( \hat{\mathbf{r}} \frac{\partial \rho}{\partial r} - \hat{\boldsymbol{\xi}} \beta_2 \frac{\partial \rho}{\partial \xi} \right) \times \hat{\boldsymbol{\phi}} + \left( \hat{\mathbf{r}} \frac{\partial \chi}{\partial r} - \hat{\boldsymbol{\xi}} \beta_2 \frac{\partial \chi}{\partial \xi} \right) \pm \hat{\boldsymbol{\xi}} \frac{1}{2r} \ , \tag{S1}$$

where $\mathbf{r}_{\text{STOV}} = (r_{\text{STOV}}, \xi_{\text{STOV}})$ is the location of the toroidal vortex core and the vortex is assumed to be embedded in a background complex field envelope $\psi = \rho e^{i\chi}$, where $\rho = \rho(r, \xi; z)$ and $\chi = \chi(r, \xi; z)$ are real. Equation (S1) gives local evolution of the STOV given values and derivatives of $\rho$ and $\chi$ at $\mathbf{r}_{\text{STOV}}$. Because STOV movement changes $\rho$ and $\chi$, they require updating at each successive $z$ position. However, use of Eq. (S1) is sufficient for roughly estimating longitudinal movement of toroidal STOVs within the pulse frame. In air, $\beta_2 = 1.5 \times 10^{-5}$ and in plasma with $N_e/N_{cr} = 0.0035$, $\beta_2 = -3.5 \times 10^{-3}$, so the derivative terms with $\beta_2$ factors in Eq. (S1) are negligible compared to the other derivative terms. The *local* evolution of $l = \pm 1$ STOV core longitudinal separation is then $k_0\,\partial\Delta\xi_{\text{STOV}}/\partial z = r^{-1}(1 + (r/\rho)\,\partial\rho/\partial r)$, with the second term negative in general. Using $z = v_g t$ and making the simplifying assumption that $\rho$ has separable $r$ and $\xi$ dependence, the rate of core separation is $\partial\Delta\xi_{\text{STOV}}/\partial t \sim v_g(\lambda/2\pi r)(1 + (r/\rho)\,\partial\rho/\partial r)$. Note that $\partial\Delta\xi_{\text{STOV}}/\partial t = 0$ when $1 + (r/\rho)\,\partial\rho/\partial r = 0$, or $\rho r = const$. In a vortex ring of radius $r$ in a fluid of density $n$, the hydrodynamic impulse, proportional to ring momentum, is $\propto nr^2$ [6]. For electromagnetic waves, $\rho^2$ plays the role of density and $\rho^2 r^2 = const.$, suggesting that STOVs move longitudinally seeking to stabilize changes in ring momentum. An upper estimate of the separation rate is $\partial\Delta\xi_{\text{STOV}}/\partial t \sim v_g(\lambda/2\pi r)$.